\documentclass[twocolumn,showpacs,pra]{revtex4}
\usepackage{amssymb}
\usepackage{amsmath}
\usepackage{graphicx}
\usepackage{dcolumn}
\usepackage{bm}

\begin{document}

\title{Evolution of entanglement for quantum mixed states}
\author{Chang-shui Yu}
\email{quaninformation@sina.com; ycs@dlut.edu.cn}
\author{X. X. Yi}
\author{He-shan Song}
\affiliation{School of Physics and Optoelectronic Technology, Dalian University of
Technology, Dalian 116024, P. R. China}
\date{\today }

\begin{abstract}
A simple relation is introduced for concurrence to describe how much the
entanglement of bipartite system is at least left if either (or both)
subsystem undergoes an arbitrary physical process. This provides a lower
bound for concurrence of mixed states (pure states are included) in contrast
to the upper bound given by Konrad et al [Nature Physics \textbf{4}, 99
(2008)]. Our results are also suitable for a general high dimensional
bipartite quantum systems.
\end{abstract}

\maketitle

\section{Introduction}

Quantum entanglement is an important physical resource in quantum
information and computation tasks such as quantum teleportation [1], quantum
key distribution [2], quantum computation [3] and so on. In realistic
quantum information processing, entanglement has to be prepared or
distributed beforehand by two distant parties in which one or more physical
systems have to be transmitted by a quantum channel. However, unlike
classical systems, quantum systems are usually fragile. It is inevitable
that environment (channel) will influence the systems of interest more or
less and induce decoherence because of the interaction with the systems, so
that entanglement is destroyed to some extent before use [4-6]. It is a key
task to evaluate the shared entanglement after the influence of environment.

In usual one has to deduce the time evolution of entanglement of the
composite system from the time evolution of the quantum state under
consideration [7], when the subsystems of the composite quantum system
undergo a physical process. That is to say, for the different potential
initial states one has to repeat the same procedure every time. Quite
recently, an important step has been taken by Konrad et al [8] who provided
an explicit evolution equation of quantum entanglement quantified by the
remarkable concurrence [9] for bipartite quantum state of qubits. It has
been shown that given any one-sided quantum channel, the concurrence of
output state corresponding to any initial pure input state of interests can
always be equivalently obtained by the product of the concurrence of input
state and that of the output state with the maximally entangled state as an
input state. However, for two-sided quantum channel or the initial mixed
states, the product of the two concurrences only provides an upper bound for
the concurrence of interests. It is obviously important to find a lower
bound of the concurrence in order to well grasp the entanglement of output
states.

In fact, quantum mechanics and quantum information processing (QIP) are not
constrained to pure states as well as $2\otimes 2$ dimensional quantum
systems. Because of imperfect experiments and inevitable disturbance of
environments, mixed (entangled) states are ubiquitous, which is inevitable
for QIP to cope with. Furthermore, on the one hand, when we face the quantum
features, especially entanglement, of meso- even macroscopic quantum
systems, such as Bose-Einstein condensates [10]. we have to deal with high
dimensional density matrices. On the other hand, it has been shown that some
QIP tasks based on high dimensional entangled states are more efficient than
those of qubits. For example, cryptographic protocols are more secure based
on quantum channel of qutrits [11-13]. Teleportation can be implemented in
faith even though a non-maximally entangled quantum channel is shared [14].
Therefore, for these large systems, it is necessary to investigate the
entanglement of high dimensional quantum systems (multi-parties included).

However, quantification of entanglement, as a precondition of studying
entanglement, is generally a hard problem which does not only lie in the
poor practicability for high dimensional quantum systems [15,16] but also
the nonlinearity on density matrices for usual entanglement measures
[17,18]. Therefore, it is often suggested to derive a lower bound to
evaluate the entanglement (Entanglement can be better evaluated, if both
upper and lower bounds are given). In this paper, we consider the evolution
of entanglement with either (or both) subsystem undergoing an arbitrary
quantum channel. \emph{With concurrence as entanglement measure, we find a
lower bound of the concurrence of any output states for a given quantum
channel based on the evolution of a probe state as input states}. It is
shown that our lower bound is not restricted to the bipartite systems of
qubits. In particular, for one-sided quantum channel, our lower bound has a
concise form. Furthermore, we also show that it is not necessary to choose
the maximally entangled state as the probe states. The paper is organized as
follows. In Sec. II, we give a lower bound for the concurrence of bipartite
quantum systems; In Sec. III, we show that the lower bound can be obtained
in terms of the initial input state and the output state with the probe
state as initial states; The conclusion is drawn finally.

\section{The lower bound for concurrence}

An $\left( N_{1}\otimes N_{2}\right) $- dimensional bipartite quantum pure
state can be written as
\begin{equation}
\left\vert \psi \right\rangle
_{AB}=\sum\limits_{i=0}^{N_{1}-1}\sum\limits_{j=0}^{N_{2}-1}\psi
_{ij}\left\vert ij\right\rangle ,
\end{equation}%
where $\left\vert ij\right\rangle $ denotes the computational basis and $%
N_{1}\times N_{2}$ matrix $\psi $ (i.e. $\left\vert \psi \right\rangle _{AB}$
without $\left\vert {}\right\rangle _{AB}$) represents the matrix notation
[19,20] of $\left\vert \psi \right\rangle _{AB}$ with matrix element $\psi
_{ij}=$ $\left\langle ij\right. \left\vert \psi \right\rangle _{AB}$ and $%
\sum\limits_{i=0}^{N_{1}-1}\sum\limits_{j=0}^{N_{2}-1}\left\vert \psi
_{ij}\right\vert ^{2}=1$. Consider the Schmidt decomposition, $\left\vert
\psi \right\rangle _{AB}$ can also be given by%
\begin{equation}
\left\vert \psi \right\rangle _{AB}=\sum\limits_{i=0}^{R-1}\lambda
_{i}\left\vert ii\right\rangle ,\sum\limits_{i}\lambda _{i}^{2}=1,
\end{equation}%
with $\lambda _{i}$ being real in decreasing order and $R=\min
\{N_{1},N_{2}\}$. The concurrence of $\left\vert \psi \right\rangle _{AB}$
can be defined [21,22] as%
\begin{eqnarray}
C(\left\vert \psi \right\rangle _{AB}) &=&\sqrt{\sum\limits_{\substack{ %
i,j=0  \\ i\neq j}}^{N_{1}-1}\sum\limits_{\substack{ p,q=0  \\ p\neq q}}%
^{N_{2}-1}\left\vert \psi _{ip}\psi _{jq}-\psi _{iq}\psi _{jp}\right\vert
^{2}} \\
&=&\sqrt{4\sum\limits_{\substack{ i,j=0  \\ i<j}}^{R-1}\lambda
_{i}^{2}\lambda _{j}^{2}}.
\end{eqnarray}%
The equivalence of eq. (3) and eq. (4) lies in that quantum pure states are
related to its Schmidt decomposition by local unitary transformations which
do not contribute to concurrence. In particular, if $N_{1}=N_{2}=2$, eq. (4)
can be reduced to $C(\left\vert \psi \right\rangle _{AB})=2\lambda
_{1}\lambda _{2}=2\left\vert \det (\psi )\right\vert $, which will be used
later. With the definitions given by eqs. (3) and (4), we obtain the
following theorem.

\textbf{Theorem 1. }For any $\left( N_{1}\otimes N_{2}\right) $- dimensional
bipartite quantum pure state $\left\vert \psi \right\rangle $,
\begin{equation}
C(\left\vert \psi \right\rangle )\geqslant \sqrt{\frac{2R}{R-1}}\left(
\underset{\left\vert \phi \right\rangle \in \mathcal{E}}{\max }\left\vert
\left\langle \psi \left\vert \phi \right. \right\rangle \right\vert ^{2}-%
\frac{1}{R}\right)
\end{equation}%
with $\mathcal{E}$ denoting the set of $\left( N_{1}\otimes N_{2}\right) $-
dimensional maximally entangled states.

\textbf{Proof}. Since $\underset{\left\vert \phi \right\rangle \in \mathcal{E%
}}{\max }\left\vert \left\langle \psi \left\vert \phi \right. \right\rangle
\right\vert ^{2}$ is not changed by any local unitary transformation on $%
\left\vert \psi \right\rangle $, $\left\vert \psi \right\rangle $ can always
be understood in the form of Schmidt decomposition, i.e., $\left\vert \psi
\right\rangle _{AB}=\sum\limits_{i=0}^{R-1}\lambda _{i}\left\vert
ii\right\rangle $. The maximally entangled state can be written as $%
\left\vert \phi \right\rangle =\frac{1}{\sqrt{R}}\sum\limits_{i=0}^{R-1}%
\left( U_{1}\otimes U_{2}\right) \left\vert ii\right\rangle $, with $U_{1}$
and $U_{2}$ denoting local unitary transformations. Suppose $a_{i}=$ $%
\left\langle ii\right. \left\vert \phi \right\rangle ,$ $a_{i}\in \lbrack 0,%
\frac{1}{\sqrt{R}}]$, then $\underset{\left\vert \phi \right\rangle \in
\mathcal{E}}{\max }\left\vert \left\langle \psi \left\vert \phi \right.
\right\rangle \right\vert ^{2}$ can be rewritten as%
\begin{gather}
\underset{\left\vert \phi \right\rangle \in \mathcal{E}}{\max }\left\vert
\left\langle \psi \left\vert \phi \right. \right\rangle \right\vert ^{2}=%
\underset{\left\vert \phi \right\rangle \in \mathcal{E}}{\max }\left\vert
\sum\limits_{i=0}^{R-1}a_{i}\lambda _{i}\right\vert ^{2}  \notag \\
\leq \underset{\left\vert \phi \right\rangle \in \mathcal{E}}{\max }\left(
\sum\limits_{i=0}^{R-1}\left\vert a_{i}\lambda _{i}\right\vert \right)
^{2}\leq \frac{1}{R}\left( \sum\limits_{i=0}^{R-1}\lambda _{i}\right) ^{2} \\
=\frac{1}{R}\left( 1+2\sum\limits_{i,j=0;i<j}^{R-1}\lambda _{i}\lambda
_{j}\right)  \notag \\
\leq \frac{1}{R}\left( 1+2\sqrt{\frac{R(R-1)}{2}\sum%
\limits_{i,j=0;i<j}^{R-1}\lambda _{i}^{2}\lambda _{j}^{2}}\right) \\
=\frac{1}{R}\left[ 1+\sqrt{\frac{R(R-1)}{2}}C(\left\vert \psi \right\rangle )%
\right] .
\end{gather}%
We arrive at the inequality (7) based on the inequality $\left(
\sum\limits_{i=1}^{n}x_{i}\right) ^{2}/n\leq \sum\limits_{i=1}^{n}x_{i}^{2}$%
. Thus from inequality (8), we can obtain
\begin{equation}
C(\left\vert \psi \right\rangle )\geqslant \sqrt{\frac{2R}{R-1}}\left(
\underset{\left\vert \phi \right\rangle \in \mathcal{E}}{\max }\left\vert
\left\langle \psi \left\vert \phi \right. \right\rangle \right\vert ^{2}-%
\frac{1}{R}\right) .
\end{equation}%
We would like to emphasize that the '=' in eq. (9) is always achieved for
the pure states of two qubits. That is to say, the right-hand side of eq.
(9) is just the concurrence of two-qubit pure states. However, for a general
high dimensional quantum systems, the '=' holds only for maximally entangled
states (in this case '=' in eq. (7) holds), which shows that the inequality
(9) provides a lower bound for concurrence in high dimension. $\hfill \Box $

Since the maximum of eq. (6) is obtained with $\left\vert a_{i}\right\vert =%
\frac{1}{\sqrt{R}}$, $\left\vert \phi \right\rangle $ can be conveniently
chosen as $\left\vert \tilde{\phi}\right\rangle $ =$\frac{1}{\sqrt{R}}%
\sum\limits_{i=0}^{R-1}\left\vert ii\right\rangle $. Therefore, a lower
bound of concurrence can be given by
\begin{equation}
C(\left\vert \psi \right\rangle )\geqslant \sqrt{\frac{2R}{R-1}}\left( \text{%
Tr}\left( \rho \left\vert \tilde{\phi}\right\rangle \left\langle \tilde{\phi}%
\right\vert \right) -\frac{1}{R}\right)
\end{equation}%
with $\rho =\left\vert \psi \right\rangle \left\langle \psi \right\vert $.
It is obvious that the lower bound in eq. (10) is less than that in eq. (9).
But it can be used conveniently because there does not exist maximization
problem.

The inequality (10) can be immediately generalized to mixed states.
Concurrence for any $\left( N_{1}\otimes N_{2}\right) $-dimensional mixed
state $\rho $ is defined as $C(\rho )=\min \sum p_{i}C(\left\vert \varphi
_{i}\right\rangle )$ where the minimum is taken over all possible
decompositions such that $\rho =\sum p_{i}\left\vert \varphi
_{i}\right\rangle \left\langle \varphi _{i}\right\vert ,\sum p_{i}=1$. Based
on the optimal decomposition $\rho =\sum q_{i}\left\vert \chi
_{i}\right\rangle \left\langle \chi _{i}\right\vert $ such that $C(\rho
)=\sum q_{i}C(\left\vert \chi _{i}\right\rangle )$, one can get%
\begin{eqnarray}
C(\rho ) &=&\sum q_{i}C(\left\vert \chi _{i}\right\rangle )  \notag \\
&\geqslant &\sqrt{\frac{2R}{R-1}}\sum q_{i}\left( \text{Tr}\left[ \left\vert
\chi _{i}\right\rangle \left\langle \chi _{i}\right\vert \left\vert \tilde{%
\phi}\right\rangle \left\langle \tilde{\phi}\right\vert \right] -\frac{1}{R}%
\right)  \notag \\
&=&\sqrt{\frac{2R}{R-1}}\left( \text{Tr}\left[ \rho \left\vert \tilde{\phi}%
\right\rangle \left\langle \tilde{\phi}\right\vert \right] -\frac{1}{R}%
\right) .
\end{eqnarray}%
Inequality (11) holds for any bipartite quantum states which provides the
key result that will be used later.

\section{Evolution of concurrence}

\subsection{One-sided quantum channel}

Next, we will show that eq. (11) can be captured by the evolution of some
probe states. Let us first consider an $(N\otimes N)$ -dimensional bipartite
quantum states $\rho $ with only one subsystem undergoing a quantum channel
represented by the superoperator \$$_{1}$, then the final state can be given
by $\rho _{f}=\frac{\left( \$_{1}\otimes \mathbf{1}\right) \rho }{p}$, where
$p=$Tr$\left[ \left( \$_{1}\otimes \mathbf{1}\right) \rho \right] $ is the
probability for channel $\$_{1}\ $which corresponds to non-trace-preserving
channel [12]. Any quantum state can be expanded in a representation spanned
by maximally entangled states given by
\begin{equation}
\left\vert \Phi _{j}\right\rangle =\frac{1}{\sqrt{N}}\sum%
\limits_{k=0}^{N-1}e^{i\frac{2j_{0}k\pi }{n}}\left\vert k\right\rangle
\left\vert k\oplus j_{1}\right\rangle ,j=Nj_{0}+j_{1},
\end{equation}%
where $j_{0},j_{1}=0,1,\cdots ,N-1,$ $\left\vert k\right\rangle $ is the
computational basis and '$\oplus $' denotes the addition modulo $N.$ A
maximally entangled state $\left\vert \Phi _{m}\right\rangle $ can also be
written as
\begin{equation}
\left\vert \Phi _{m}\right\rangle =\left( \Phi _{m}P^{-1}\otimes \mathbf{1}%
\right) \left\vert P\right\rangle ,
\end{equation}%
where $\left\vert P\right\rangle
=\sum\limits_{i,j=0}^{N-1}a_{ij}\left\vert ij\right\rangle $, called
'probe quantum state' in this paper, is a generic entangled pure
state with full-rank $P$ (which can be explicitly written as $$
P=\left(
\begin{array}{cccc}
a_{00} & a_{01} & \cdots  & a_{0\left( N-1\right) } \\
a_{10} & a_{11} & \cdots  & a_{1\left( N-1\right) } \\
\vdots  & \vdots  & \ddots  & \vdots  \\
a_{\left( N-1\right) 0} & a_{\left( N-1\right) 1} & \cdots  & a_{\left(
N-1\right) \left( N-1\right) }%
\end{array}%
\right),$$ or can be directly obtained by the method provided below
eq. (1).) and $P^{-1}$  denotes the inverse matrix of $P$. $\Phi
_{m} $ are simple unitary transformations determined by eq. (12).
For example, for the state of a pair of qubits, $\Phi _{m}$
correspond to the three Pauli matrices and the identity,
respectively. Since any quantum state $\left\vert
\psi \right\rangle $ can be given [19], based on maximally entangled state $%
\left\vert \tilde{\phi}\right\rangle ,$ by%
\begin{equation}
\left\vert \psi \right\rangle =\sqrt{R}\left( \psi \otimes \mathbf{1}\right)
\left\vert \tilde{\phi}\right\rangle =\sqrt{R}\left( \mathbf{1}\otimes \psi
^{T}\right) \left\vert \tilde{\phi}\right\rangle ,
\end{equation}%
$\left\vert \psi \right\rangle $ can also be written as
\begin{equation}
\left\vert \psi \right\rangle =\left( \psi P^{-1}\otimes \mathbf{1}\right)
\left\vert P\right\rangle =\left( \mathbf{1}\otimes \psi ^{T}\left(
P^{-1}\right) ^{T}\right) \left\vert P\right\rangle ,
\end{equation}%
where the superscript $T$ denotes transpose. Hence, we have%
\begin{gather}
Tr\left[ \rho _{f}\left\vert \tilde{\phi}\right\rangle \left\langle \tilde{%
\phi}\right\vert \right] =\frac{1}{p}\text{Tr}\left[ \left( \$_{1}\otimes
\mathbf{1}\right) \rho \left\vert \tilde{\phi}\right\rangle \left\langle
\tilde{\phi}\right\vert \right]   \notag \\
=\frac{1}{p}\text{Tr}\left[ \rho \left( \$_{1}^{\dagger }\otimes \mathbf{1}%
\right) \left( \left\vert \tilde{\phi}\right\rangle \left\langle \tilde{\phi}%
\right\vert \right) \right]   \notag \\
=\frac{1}{p}\text{Tr}\left[ S\rho ^{\ast }S\left( \$_{1}\otimes \mathbf{1}%
\right) \left( \left\vert \tilde{\phi}\right\rangle \left\langle \tilde{\phi}%
\right\vert \right) \right] ,
\end{gather}%
where $S$ is the swapping operator defined as $S\left\vert j\right\rangle
\left\vert k\right\rangle =\left\vert k\right\rangle \left\vert
j\right\rangle $ and we apply eq. (14) and $S\left\vert \tilde{\phi}%
\right\rangle =\left\vert \tilde{\phi}\right\rangle $ to the third '='. Eq.
(16) can be understood by the Kraus representation of superoperator $\$_{1}$
[23]. Based on eq. (13) and eq. (15), eq. (16) can arrive at
\begin{gather}
Tr\left[ \rho _{f}\left\vert \tilde{\phi}\right\rangle \left\langle \tilde{%
\phi}\right\vert \right]   \notag \\
=\frac{1}{p}\text{Tr}S\rho ^{\ast }S\left( \$_{1}\otimes \mathbf{1}\right) %
\left[ \left( \mathbf{1}\otimes \tilde{\phi}^{T}\left( P^{-1}\right)
^{T}\right) \left\vert P\right\rangle \left\langle P\right\vert \mathbf{1}%
\otimes \left( P^{-1}\right) ^{\ast }\tilde{\phi}^{\ast }\right]   \notag \\
=\frac{1}{p_{t}R}\text{Tr}S\rho ^{\ast }S\left[ \mathbf{1\otimes }\left(
P^{-1}\right) ^{T}\right] \left[ \frac{\left( \$_{1}\otimes \mathbf{1}%
\right) \left\vert P\right\rangle \left\langle P\right\vert }{p^{\prime }}%
\right] \left[ \mathbf{1\otimes }\left( P^{-1}\right) ^{\ast }\right] ,
\end{gather}%
where $p^{\prime }=$Tr$\left( \$_{1}\otimes \mathbf{1}\right) \left\vert
P\right\rangle \left\langle P\right\vert $, $p_{t}=p/p^{\prime }$ and the
star denotes conjugate operation. An alternative derivation can be done by
substituting $\rho =\sum p_{i}\left\vert \varphi _{i}\right\rangle
\left\langle \varphi _{i}\right\vert $ into the first line of eq. (16). In a
representation of maximally entangled states, we can have%
\begin{eqnarray}
p_{t} &=&\text{Tr}\left[ \left( \$_{1}\otimes \mathbf{1}\right) \rho \right]
/p^{\prime }  \notag \\
&=&\text{Tr}\sum\limits_{m}\left[ \left( \$_{1}\otimes \mathbf{1}\right)
\rho \right] \left\vert \Phi _{m}\right\rangle \left\langle \Phi
_{m}\right\vert /p^{\prime }  \notag \\
&=&\text{Tr}\sum\limits_{m}\left[ \frac{\left( \$_{1}\otimes \mathbf{1}%
\right) \left\vert P\right\rangle \left\langle P\right\vert }{p^{\prime }}%
\right]   \notag \\
&&\cdot \left[ \Phi _{m}\otimes \left( P^{-1}\right) ^{\ast }\right] S\rho
^{\ast }S\left[ \Phi _{m}^{\dagger }\otimes \left( P^{-1}\right) ^{T}\right]
.
\end{eqnarray}%
In fact, if the reduced density of the initial state $\rho $ is considered, $%
p_{t}$ has an alternative and concise form. Let $\rho _{A}=$Tr$_{B}\rho
_{AB}=$Tr$_{B}\rho $ with Tr$_{B}$ denoting trace over subsystem B. Consider
a decomposition of $\rho =\sum p_{i}\left\vert \varphi _{i}\right\rangle
\left\langle \varphi _{i}\right\vert $, $\rho _{A}$ can be rewritten as $%
\rho _{A}=\sum p_{i}\varphi _{i}\varphi _{i}^{\dag }$. Thus $p_{t}$ can also
be given by
\begin{equation}
p_{t}=\text{Tr}\left[ \frac{\left( \$_{1}\otimes \mathbf{1}\right)
\left\vert P\right\rangle \left\langle P\right\vert }{p^{\prime }}\right] %
\left[ \mathbf{1\otimes }\left( P^{-1}\rho _{A}\left[ P^{-1}\right] ^{\dag
}\right) ^{\ast }\right] .
\end{equation}%
Eq. (19) has a concise form without summation. Substitute eq. (18) or eq.
(19) into eq. (17), one can find that $Tr\left[ \rho _{f}\left\vert \tilde{%
\phi}\right\rangle \left\langle \tilde{\phi}\right\vert \right] $ has been
given by a simple algebra on the evolution of the probe state $\left\vert
P\right\rangle $ and the original density matrix. That is to say, the lower
bound of concurrence in eq. (11) can be captured by the evolution of the
given probe state which can be formally written as
\begin{gather}
C\left[ \left( \$_{1}\otimes \mathbf{1}\right) \rho \right]   \notag \\
\geqslant \sqrt{\frac{2R}{R-1}}\left( \text{Tr}\left[ f\left( \frac{\left(
\$_{1}\otimes \mathbf{1}\right) \left\vert P\right\rangle \left\langle
P\right\vert }{p^{\prime }}\right) \rho ^{\ast }\right] -\frac{1}{R}\right) ,
\end{gather}%
where $f\left( x\right) =\frac{1}{p_{t}R}S\left[ \mathbf{1\otimes }\left(
P^{-1}\right) ^{T}\right] \left[ x\right] \left[ \mathbf{1\otimes }\left(
P^{-1}\right) ^{\ast }\right] S.$ It is obvious that the lower bound of
concurrence is determined by the evoluted probe state.

\subsection{Two-sided quantum channel}

Eq. (20) can immediately be generalized to the case of two-sided quantum
channel, however the form might not be as simple as eq. (17). Since the
lower bound given in eq. (17) is valid for mixed initial states, the lower
bound for two-sided quantum channel can be easily\ obtained by replacing $%
\rho $ in eq. (17) by $(\mathbf{1}\otimes \$_{2})\rho $. Thus Tr$\left[
\left\vert \tilde{\phi}\right\rangle \left\langle \tilde{\phi}\right\vert
\left( \$_{1}\otimes \$_{2}\right) \rho \right] $ can be written as%
\begin{gather}
\text{Tr}\left[ \left\vert \tilde{\phi}\right\rangle \left\langle \tilde{\phi%
}\right\vert \left( \$_{1}\otimes \$_{2}\right) \rho \right] =Tr\left[
\left\vert \tilde{\phi}\right\rangle \left\langle \tilde{\phi}\right\vert
\left( \$_{1}\otimes \mathbf{1}\right) \left( \mathbf{1}\otimes
\$_{2}\right) \rho \right]  \notag \\
=\frac{1}{R}\text{Tr}\sum\limits_{k}\left\{ \left[ \mathbf{1\otimes }\left(
P^{-1}\right) ^{T}\right] \left[ \left( \$_{1}\otimes \mathbf{1}\right)
\left\vert P\right\rangle \left\langle P\right\vert \right] \left[ \mathbf{%
1\otimes }\left( P^{-1}\right) ^{\ast }\right] \right.  \notag \\
\times \left. S\left[ (\mathbf{1}\otimes \$_{2})\left\vert \Psi _{\rho
k}\right\rangle \left\langle \Psi _{\rho k}\right\vert \right] ^{\ast
}S\right\} ,
\end{gather}%
where we have replaced $\rho $ by a potential decomposition of $\rho
=\sum\limits_{k}\left\vert \Psi _{\rho k}\right\rangle \left\langle \Psi
_{\rho k}\right\vert $ which is especially referred to the eigenvalue
decomposition for simplicity. Applying eq. (15) to $\left\vert \Psi _{\rho
k}\right\rangle $, eq. (21) leads to
\begin{gather}
\text{Tr}\left[ \left\vert \tilde{\phi}\right\rangle \left\langle \tilde{\phi%
}\right\vert \left( \$_{1}\otimes \$_{2}\right) \rho \right]  \notag \\
=\frac{1}{R}\text{Tr}\sum\limits_{k}\left\{ \left[ \mathbf{1\otimes }\left(
P^{-1}\right) ^{T}\right] \left[ \left( \$_{1}\otimes \mathbf{1}\right)
\left\vert P\right\rangle \left\langle P\right\vert \right] \left[ \mathbf{%
1\otimes }\left( P^{-1}\right) ^{\ast }\right] \right.  \notag \\
\times \left. S\left( \Psi _{\rho k}P^{-1}\otimes \mathbf{1}\right) ^{\ast }%
\left[ (\mathbf{1}\otimes \$_{2})\left\vert P\right\rangle \left\langle
P\right\vert \right] ^{\ast }\left( \left[ P^{-1}\right] ^{\dag }\Psi _{\rho
k}^{\dag }\otimes \mathbf{1}\right) ^{\ast }S\right\}  \notag \\
=\frac{1}{R}\text{Tr}\sum\limits_{k}\left\{ \left[ \left( \$_{1}\otimes
\mathbf{1}\right) \left\vert P\right\rangle \left\langle P\right\vert \right]
^{\ast }\left( \mathbf{1}\otimes P^{-1}\Psi _{\rho k}P^{-1}\right) \right.
\notag \\
\times \left. S\left[ (\mathbf{1}\otimes \$_{2})\left\vert P\right\rangle
\left\langle P\right\vert \right] S\left( \mathbf{1}\otimes \left(
P^{-1}\Psi _{\rho k}P^{-1}\right) ^{\dag }\right) \right\} .
\end{gather}%
\newline
Substituting eq. (21) into eq. (17), one can find that the lower bound of
concurrence $C\left[ \left( \$_{1}\otimes \$_{2}\right) \rho \right] $ can
be captured by the evolutions of the probe state under the two quantum
channels. In order to avoid the decomposition of the initial state $\rho $,
one can expand $\left[ (\mathbf{1}\otimes \$_{2})\left\vert P\right\rangle
\left\langle P\right\vert \right] $ in the representation of maximally
entangled states. Thus eq. (22) can be rewritten as
\begin{gather}
\text{Tr}\left[ \left\vert \tilde{\phi}\right\rangle \left\langle \tilde{\phi%
}\right\vert \left( \$_{1}\otimes \$_{2}\right) \rho \right]  \notag \\
=\frac{1}{R}\text{Tr}\sum\limits_{mnk}\left\langle \Phi _{m}\right\vert %
\left[ (\mathbf{1}\otimes \$_{2})\left\vert P\right\rangle \left\langle
P\right\vert \right] \left\vert \Phi _{n}\right\rangle \cdot \left[ \left(
\$_{1}\otimes \mathbf{1}\right) \left\vert P\right\rangle \left\langle
P\right\vert \right] ^{\ast }  \notag \\
\times \left[ \left( \mathbf{1}\otimes P^{-1}\Psi _{\rho k}P^{-1}\right) %
\right] S\left\vert \Phi _{m}\right\rangle \left\langle \Phi _{n}\right\vert
S\left[ \mathbf{1}\otimes \left( P^{-1}\Psi _{\rho k}P^{-1}\right) ^{\dag }%
\right]  \notag \\
=\frac{1}{R}\text{Tr}\sum\limits_{mn}\left\langle \Phi _{m}\right\vert \left[
(\mathbf{1}\otimes \$_{2})\left\vert P\right\rangle \left\langle
P\right\vert \right] \left\vert \Phi _{n}\right\rangle \cdot \left[ \left(
\$_{1}\otimes \mathbf{1}\right) \left\vert P\right\rangle \left\langle
P\right\vert \right] ^{\ast }  \notag \\
\times \left[ \left( \Phi _{m}^{T}\left[ P^{-1}\right] ^{T}\otimes
P^{-1}\right) \right] S\rho S\left[ \left[ P^{-1}\right] ^{\ast }\Phi
_{n}^{\ast }\otimes \left( P^{-1}\right) ^{\dag }\right] .
\end{gather}%
Finally, it is worth noting that maximally entangled state is a special
choice of our probe states. Furthermore, the value of the lower bound of
concurrence does not depend on the choice of probe state. It is obvious that
eq. (17) and eq. (18) correspond to the trace-preserving quantum channels.
The general results for non-trace-preserving channels are omitted here,
which can be directly given by adding some normalization constants like the
case of one-sided quantum channel.

For integrity, we show that the upper bound given in Ref. [8] can be
captured by the given probe state $\left\vert P\right\rangle $, but the
value of the bound is not changed. Suppose that $\left\vert \psi
\right\rangle $ is a bipartite quantum state of qubits, then the concurrence
can be given by%
\begin{equation}
C(\left\vert \psi \right\rangle )=2\left\vert \det \left( \psi \right)
\right\vert .
\end{equation}%
If one of the subsystems undergoes a quantum channel $\$_{1}$, the final
state can be given by $\rho _{f}=\left( \$_{1}\otimes \mathbf{1}\right)
\left\vert \psi \right\rangle \left\langle \psi \right\vert /p_{1}$, with $%
p_{1}=$Tr$\left( \$_{1}\otimes \mathbf{1}\right) \left\vert \psi
\right\rangle \left\langle \psi \right\vert $. Thus we have
\begin{eqnarray}
&&\rho _{f}\left( \sigma _{y}\otimes \sigma _{y}\right) \rho _{f}^{\ast
}\left( \sigma _{y}\otimes \sigma _{y}\right)  \notag \\
&=&\left[ \frac{\det \left( \psi \right) }{p_{1}p_{2}\det \left( P\right) }%
\right] ^{2}\rho _{P}\left( \sigma _{y}\otimes \sigma _{y}\right) \rho
_{P}^{\ast }\left( \sigma _{y}\otimes \sigma _{y}\right) ,
\end{eqnarray}%
with $\rho _{P}=\left( \$_{1}\otimes \mathbf{1}\right) \left\vert
P\right\rangle \left\langle P\right\vert /p_{2}\ $and $p_{2}=$Tr$\left(
\$_{1}\otimes \mathbf{1}\right) \left\vert P\right\rangle \left\langle
P\right\vert $. Based on eq. (25), we can obtain
\begin{equation}
C(\rho _{f})=\left\vert \frac{\det \left( \psi \right) }{\det \left(
P\right) }\right\vert C(\rho _{P})=\frac{C(\left\vert \psi \right\rangle
)\cdot C(\rho _{P})}{2\left\vert \det \left( P\right) \right\vert }.
\end{equation}%
It is obvious for a mixed initial state $\varrho $ that eq. (25) is extended
to
\begin{equation}
C(\varrho _{f})\leq \frac{C(\varrho )\cdot C(\rho _{P})}{2\left\vert \det
\left( P\right) \right\vert }.
\end{equation}%
For two-sided quantum channel, one can easily obtain
\begin{equation}
C(\varrho _{f})\leq \frac{C(\varrho )\cdot C(\rho _{P1})}{2\left\vert \det
\left( P\right) \right\vert }\cdot \frac{C(\rho _{P2})}{2\left\vert \det
\left( P\right) \right\vert },
\end{equation}%
with $\rho _{P1}=\left( \$_{1}\otimes \mathbf{1}\right) \left\vert
P\right\rangle \left\langle P\right\vert /p_{2}$, $\rho _{P2}=\left( \mathbf{%
1}\otimes \$_{2}\right) \left\vert P\right\rangle \left\langle P\right\vert
/p_{2}^{\prime }$ and $p_{2}^{\prime }=$Tr$\left( \mathbf{1}\otimes
\$_{2}\right) \left\vert P\right\rangle \left\langle P\right\vert .$
\begin{figure}[tbp]
\includegraphics[width=9.0cm]{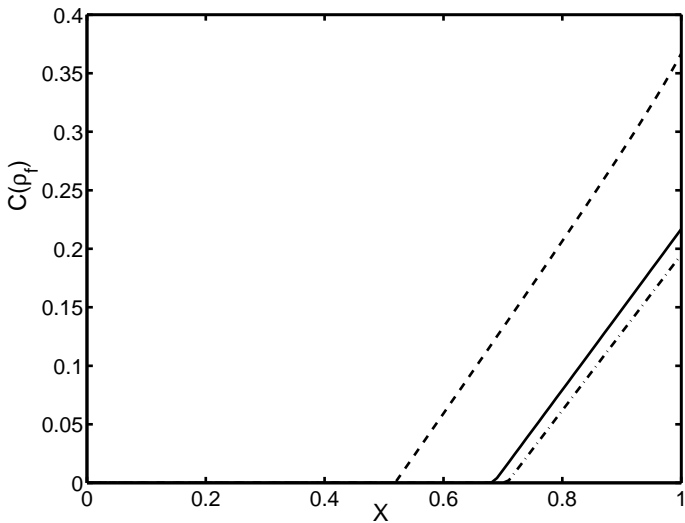} % Here is how to import EPS art
\caption{(Dimensionless)The upper bound of concurrence (dashed line)
given in Ref. [12], the concurrence (solid line) and the lower bound
(dash-dot line) of quantum state $\protect\rho _{f}$ vs $x$.
$\protect\rho _{f}$ is the final output state of an initial state
$\protect\rho $ with the subsystems undergoing a quantum channel
$\$_{1}$ and $\$_{2}$, respectively.}
\end{figure}

Thus concurrence (especially bipartite concurrence of qubits) can be
better evaluated by the lower bound and the upper bound given in eq.
(28) than by only one bound. From eq. (20) as well as eq. (22), one
might think that it is not so convenient compared with the upper
bound in Ref. [8], because the upper bound is a simple linear
relationship between the concurrence of evolved probe state and that
of the initial states. However, generally speaking, concurrence per
se is not a direct observable [24-26], therefore, to evaluate
concurrence of a state in practice, one has to evaluate the quantum
state by quantum state tomography [27] and then turns to a
mathematical procedure. In other words, it is inevitable for Ref.
[8] to evaluate quantum states in practical scenario. In this sense,
we think that their practicability is almost the same. What is more,
our lower bound has obvious advantages: 1) The lower bound has the
consistent spirit with entanglement measures of mixed states for
which the lower bounds (infimum) are usually needed; 2) Bounds,
especially lower bounds, with elegant forms like Ref. [8] might be
difficult to provided. In particular, so far there have not been
analytic results of entanglement measure (in particular, concurrence
included) for general high-dimensional quantum systems, therefore,
if the bounds of entanglement for high-dimensional mixed states
still include the calculation of high dimensional entanglement
measures, it only formally provides an elegant relationship, but it
has usually poor practicability. 3) The derivation based on our
extended lower bound provides a universal method for all analytic
bounds of entanglement measure, with which one can only focus all
the attention on the tightness of the bounds, but there might be a
great deal of difference on the complexity of the final results
between different lower bounds.

\section{A simple application}

As an application, we only consider a bipartite system of qubits, because
there exist analytic concurrence and acceptable upper bounds for bipartite
mixed state of qubits. Thus one can directly find the tightness of our lower
bound by comparing it with the concurrence and the upper bound. The
bipartite quantum state we considered is given by
\begin{equation}
\rho =x\rho _{r}+\frac{(1-x)}{4}\mathbf{1},x\in \lbrack 0,1]
\end{equation}%
with
\begin{equation}
\rho _{r}=\left(
\begin{array}{cccc}
0.4322 & 0.2113 & 0.1073 & 0.3369 \\
0.2113 & 0.1845 & 0.0406 & 0.1798 \\
0.1073 & 0.0406 & 0.0504 & 0.1144 \\
0.3369 & 0.1798 & 0.1144 & 0.3330%
\end{array}%
\right)
\end{equation}%
randomly generated by \textbf{Matlab 6.5}.\textbf{\ }We suppose that each
subsystem of $\rho $ undergoes an amplitude-damping quantum channel given in
Kraus representation [28] as
\begin{equation}
\$_{1}:M_{1}=\left(
\begin{array}{cc}
1 & 0 \\
0 & \sqrt{0.8}%
\end{array}%
\right) ,M_{2}=\left(
\begin{array}{cc}
0 & \sqrt{0.2} \\
0 & 0%
\end{array}%
\right) ;
\end{equation}%
\begin{equation}
\$_{2}:\tilde{M}_{1}=\left(
\begin{array}{cc}
1 & 0 \\
0 & \sqrt{0.7}%
\end{array}%
\right) ,\tilde{M}_{2}=\left(
\begin{array}{cc}
0 & \sqrt{0.3} \\
0 & 0%
\end{array}%
\right) .
\end{equation}%
The final state can be written as
\begin{equation}
\rho _{f}=\left( \$_{1}\otimes \$_{2}\right) \rho .
\end{equation}%
The upper bound, the concurrence itself and the lower bound are given in
FIG. 1, respectively, from which we can find that our lower bound is a good
evaluation of concurrence. The probe state can be chosen freely.

\section{Conclusion and Discussion}

We have presented a lower bound of concurrence. In particular, when the
quantum state under consideration evolves under a quantum channel, the lower
bound of concurrence can be completely captured by the evolution of probe
states. Thus, the evolution of concurrence for any initial state can be well
evaluated by the lower bound and the upper bound. Furthermore, the lower
bound is also suitable for high-dimensional quantum state. We would like to
emphasize that, even though the form of the lower bound seems not to be as
elegant as that of the upper bound, but their practicality is almost the
same.

\section{Acknowledgements}

This work was supported by the National Natural Science Foundation
of China, under Grant No. 10805007 and No. 10875020, and the
Doctoral Startup Foundation of Liaoning Province.

\end{document}